\documentclass[11pt,twoside]{article}

\usepackage{asp2004}
\usepackage{epsf}
\usepackage{psfig}
\usepackage{lscape}

\begin{document}
\title{Hard X-ray and Infrared Emission from Apparently Single White Dwarfs}
\author{Y.-H.\ Chu$^1$, R.~A.\ Gruendl$^1$, M.~A.\ Guerrero$^2$, and 
K.~Y.-L.\ Su$^3$}
\affil{$^1$ Astronomy Department, University of Illinois, 1002 W. Green St., Urbana,
IL 61801, USA\\
$^2$ Instituto de Astrof\'{\i}sica de Andaluc\'{\i}a (IAA, CSIC),
Camino Bajo de Hu\'etor, 50, 18008 Granada, Spain \\
$^3$ Steward Observatory, University of Arizona, 933 N. Cherry Ave.,
Tucson, AZ 85721, USA}

\begin{abstract} 
The photospheric emission of a white dwarf (WD) is not expected to
be detectable in hard X-rays or the mid-IR.  Hard X-ray ($\sim$1 keV)
emission associated with a WD is usually attributed to a binary companion;
however, emission at 1 keV has been detected from three WDs without 
companions: KPD\,0005+5106, PG\,1159, and WD\,2226$-$210.  
The origin of their hard X-ray emission is unknown, although it
has been suggested that WD\,2226$-$210 has a late-type companion
whose coronal activity is responsible for the hard X-rays.
Recent {\it Spitzer} observations of WD\,2226$-$210 revealed mid-IR
excess emission indicative of the existence of a dust disk.
It now becomes much less clear whether WD\,2226$-$210's hard X-ray 
emission originates from the corona of a late-type companion or from
the accretion of the disk material. High-quality X-ray observations
and mid-IR observations of KPD\,0005+5106 and PG\,1159 are needed
to help us understand the origin of their hard X-ray emission.
\end{abstract}



\section{Hard X-ray Emission from Apparently Single White Dwarfs}

White dwarfs (WDs) can be associated with three types of X-ray sources:
(1) photospheric X-ray emission from the WD itself, (2) accretion of material
from a close binary companion, as in cataclysmic variables, and (3) coronal
X-ray emission from a late-type binary companion, such as dMe stars.  The
latter two types of sources originate from WDs in binary systems and 
commonly show X-ray emission at 1 keV.  In contrast, the photospheric
X-ray emission from WDs is much softer, well below 0.5 keV.  No hard X-ray
emission is expected from single WDs.

Inspired by the hard X-ray emission from WD\,2226$-$210, the central star 
of the Helix Nebula \citep{Guetal01}, we made a systematic search for
hard X-ray emission from WDs using the WD catalog compiled by \citet{MS99}
and the Second {\it ROSAT} Source Catalogs of Pointed Observations with 
the PSPC with and without the Boron Filter (2RXF and 2RXP).  Among the 
124 X-ray sources that are found coincident with WDs, 22 show hard X-ray
emission peaking near 0.8-0.9 keV.  After identifying binary systems and
extraneous X-ray sources, we find two WDs, KPD\,0005+5106
(= WD\,0005+511) and PG\,1159 (= WD\,1159$-$034), that are associated 
with hard X-ray emission and appear to be single \citep{Oetal03,Cetal04a}.

\begin{figure}[ht]
\centerline{
\psfig{file=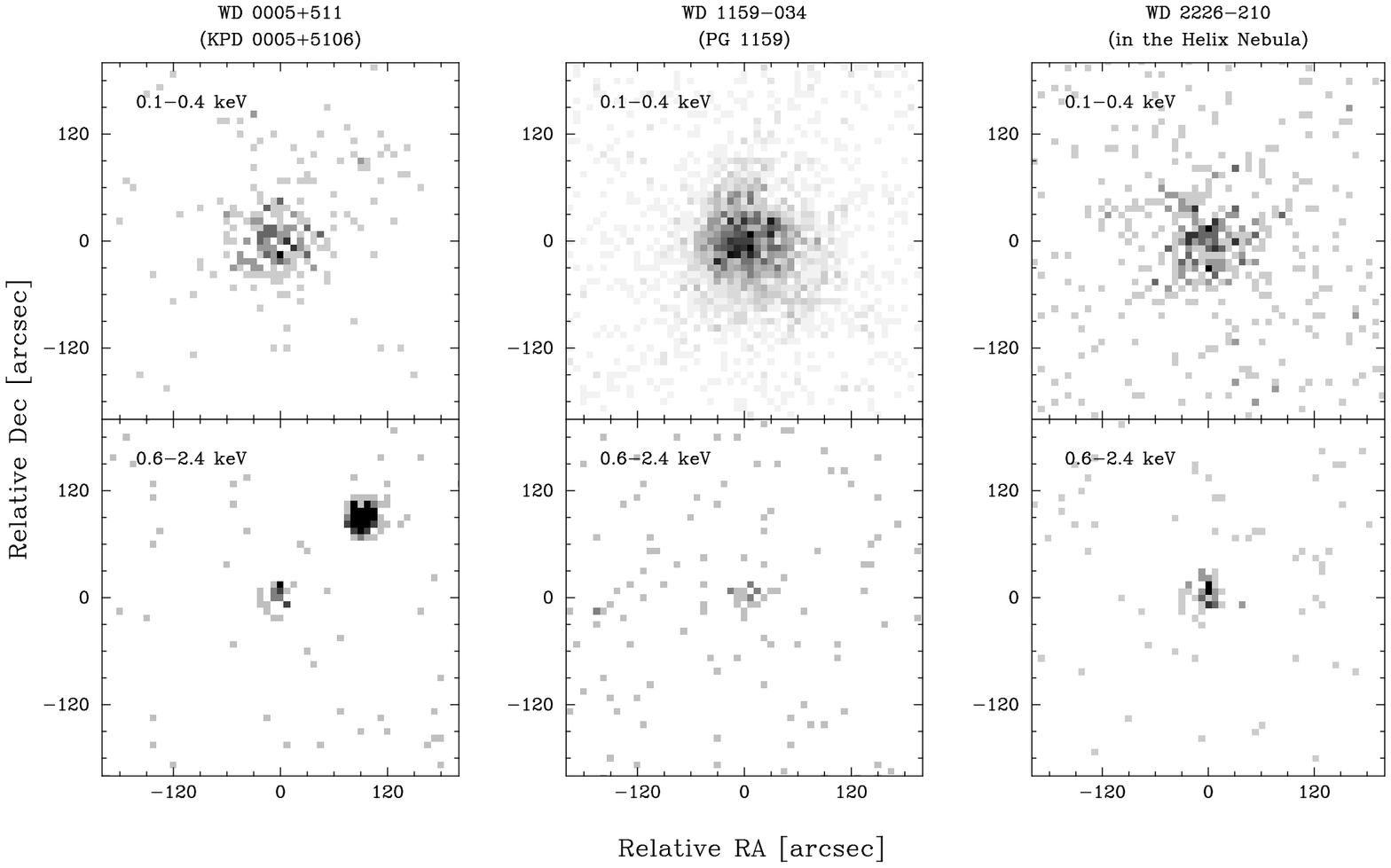,width=5in,angle=-90}
}
\vspace*{0.3cm}
\centerline{
\psfig{file=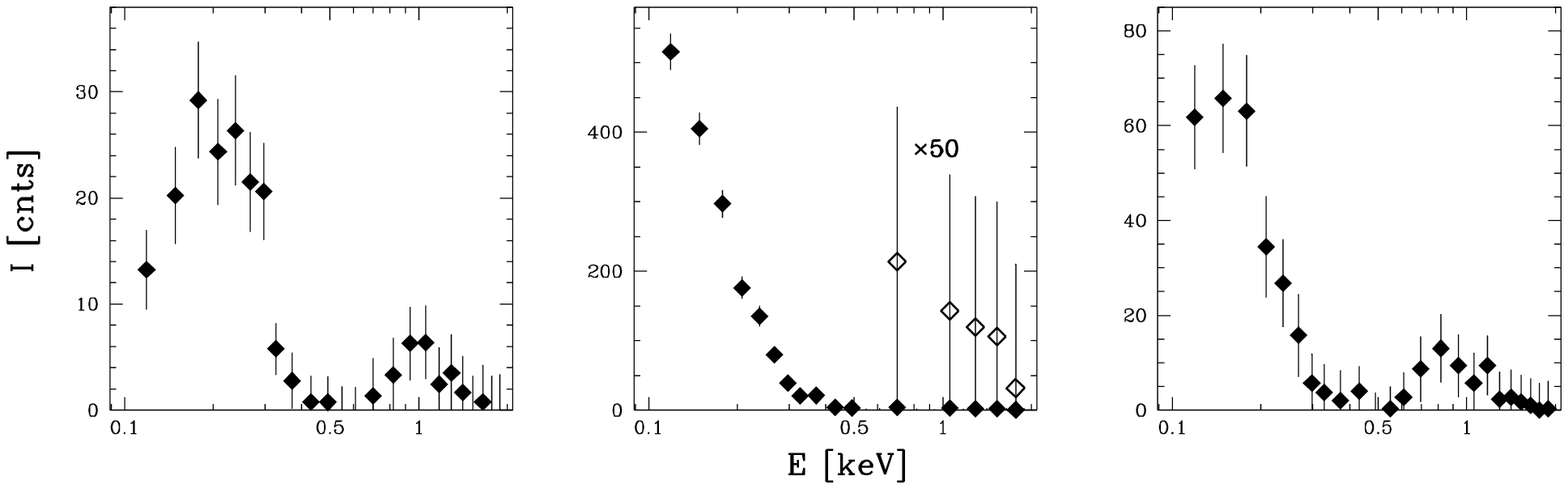,width=5.in,angle=0}
}
\caption{
{\it ROSAT} PSPC observations of KPD\,0005+5106, PG\,1159, and
WD\,2226$-$210. The top two rows show soft (0.1--0.4 keV)
and hard (0.6--2.4 keV) band images.  The poor point-spread-function
in the soft band is caused by an electronic ghost image at energies
below 0.2 keV.
The bottom row shows the PSPC spectra.  To show the hard counts of
PG\,1159, its spectrum above 0.6 keV is scaled up by a factor of 50
as marked and plotted in open symbols.
}
\end{figure}

The hard X-ray emission from these three apparently single WDs is
illustrated in Figure 1.  The {\it ROSAT} Position Sensitive 
Proportional Counter (PSPC) images in the 0.6--2.4 keV
energy band show clearly point sources coincident with the WDs, and
the PSPC spectra show distinct emission near 1 keV.

\subsection{KPD\,0005+5106}

KPD\,0005+5106 was detected in the {\it ROSAT} All Sky Survey,
and the analysis of its soft X-ray emission led to the suggestion
of a (2--3)$\times$10$^5$ K corona \citep{Fetal93}.  The 1 keV
hard X-ray emission was detected in a pointed {\it ROSAT} PSPC
observation made with the boron filter \citep{Oetal03}.
KPD\,0005+5106 has been observed with the {\it Chandra} Low
Energy Transmission Grating Spectrometer.  This observation
detected only the soft X-ray emission, and the spectrum can be
described well by non-LTE photospheric models that contain some 
Fe, but the models cannot reproduce the 1 keV emission detected
by {\it ROSAT} \citep{DW05}.

KPD\,0005+5106 does not have any known companion.  Its photometric
measurements ($V$=13.32, $J$=13.93, $H$=14.13, $K$=14.18) 
do not show an IR excess \citep{Oetal03}.  The lack of IR excess
and the presence of hard X-ray emission together place even more 
stringent constraints against the existence of a late-type companion
\citep{Cetal04b}.
The {\it ROSAT} PSPC count rate, 0.005$\pm$0.001 counts s$^{-1}$ 
in the 0.4--2.0 keV band with the boron filter is estimated to 
correspond to roughly an X-ray luminosity of $L_{\rm X}$ = 
(0.4--4)$\times$10$^{30}$ ergs s$^{-1}$.  
If this hard X-ray emission originates from the corona of a 
late-type companion with a canonical 
$L_{\rm X}/L_{\rm bol}$ = 10$^{-3}$--10$^{-4}$, this companion
should have $K$=11--12, but this is much brighter than the
observed $K$=14.18.  Therefore, the hard X-rays of KPD\,0005+5106
cannot originate from the corona of a late-type companion.

KPD\,0005+5106 had numerous high-dispersion spectroscopic observations.
Using {\it IUE} observations, \citet{Hetal98} reported average velocities
of the photospheric N\,V and C\,IV lines (36.2 km~s$^{-1}$), circumstellar
N\,V, Si\,IV, and C\,IV lines ($-$6.2 km~s$^{-1}$), and interstellar
N\,I, C\,II, Si\,II, and S\,II lines ($-$7.5 km~s$^{-1}$).
Recently, using {\it FUSE} observations, \citet{Otte04} discovered
an O\,VI-emitting nebula around KPD\,0005+5106.  \citet{Cetal04b} 
have analyzed the nebular environment of KPD\,0005+5106, using
radial velocities of all lines and spatial distributions of 
the emission lines.  They show that KPD\,0005+5106 resides in
a photoionized interstellar H\,II region with a density of 
$\sim0.8$ H-atom cm$^{-3}$.  This interstellar H\,II region
is responsible for the high-ionization ``circumstellar" lines 
reported by \citet{Hetal98}.  Therefore, there is no evidence 
of a circumstellar nebula or wind outflow from KPD\,0005+5106.

While the {\it ROSAT} PSPC resolution cannot exclude the possibility
of a chance superposition of a background AGN X-ray source near 
KPD\,0005+5106, the stellar O\,VIII emission \citep{Wetal96,Setal97}
places a strong constraint on the location of the hard X-ray source.
The O\,VIII line is emitted from recombinations of O$^{+8}$.
As the excitation potential of O$^{+8}$ is 871 eV, energetic
photons are needed; thus, the hard X-ray source must be local to
KPD\,0005+5106.  The origin of the hard X-ray emission from
KPD\,0005+5106 is most puzzling.

\subsection{PG\,1159}

PG\,1159 is a bright soft X-ray source.  Its hard X-ray emission
was detected in two independent {\it ROSAT} PSPC observations, 
RP701202N00 (13.6 ks) without the boron filter and RF200430N00
(9.8 ks) with the boron filter.  Each observation detected 
12$\pm$3 counts in the 0.6--2.4 keV band.  

PG\,1159 does not have any known companion.  Its photometric
measurements ($V$=14.89, $J$=15.58, $H$=15.87, $K$=15.78)
do not show an IR excess \citep{Oetal03}.  PG\,1159
is a pulsator; therefore, a stellar companion can be definitively
ruled out by high-precision measurements of its pulsation modes
\citep{KB94}.

Both the presence of hard X-ray emission and the single-star
status of PG\,1159 are well established.  The point spread function
of the {\it ROSAT} PSPC has a FWHM of 30$''$--40$''$.  Both the
spectral shape and the spatial coincidence between the hard X-ray 
source and PG\,1159 can be improved by a deep {\it Chandra} or 
{\it XMM-Newton} observation.

\subsection{WD\,2226$-$210}

WD\,2226$-$210 is the central star of the Helix Nebula.
{\it Chandra} ACIS-S observations show a point X-ray source 
coincident with the star within 0\farcs5.  The ACIS spectrum
can be fitted by a thin plasma emission model with a temperature
of (7--8)$\times$10$^6$ K and an X-ray luminosity of $L_{\rm X}$
= 3$\times$10$^{29}$ ergs~s$^{-1}$ in the 0.3--2.0 keV range.
These X-ray properties and the 25\% decline of the X-ray
luminosity during the period of {\it Chandra} observation
prompted \citet{Guetal01} to suggest the existence of
a dMe companion.  This hypothesis is supported by the
observed variability of the stellar H$\alpha$ emission 
line profile \citep{Gretal01}.

An {\it HST} search for companions of WD\,2226$-$210 has yielded 
null results, indicating that no companion earlier than M5 exists
\citep{Cetal99}.  If there is a late type companion, its luminosity
has to be lower than 4.2$\times$10$^{31}$ ergs~s$^{-1}$.  If this
hypothetical companion is responsible for the hard X-ray emission,
its $L_{\rm X}/L_{\rm bol}$ is $\ge$7$\times$10$^{-3}$, much 
higher than the canonical 10$^{-4}$--10$^{-3}$.
It is uncertain whether WD\,2226$-$210 indeed has a late-type
dMe companion.  It is also puzzling how a late-type dMe star
can have such a high $L_{\rm X}/L_{\rm bol}$.

\section{Mid-Infrared Excess of WD\,2226$-$210}

Excess emission at near-IR wavelengths is often used
to diagnose the presence of a late-type companion.
The Helix central star has no near-IR excess; however, a 
mid-IR excess is revealed by observations made with the
{\it Spitzer Space Telescope} IRAC at 3.6, 4.5, 5.8, and 
8.0 $\mu$m and MIPS at 24, 70, and 160 $\mu$m \citep{Suetal06}.
The central star, WD\,2226$-$210, is detected at all 
except the 160 $\mu$m MIPS band. 
Figure 2 presents the {\it Spitzer} images of the Helix
at 3.6, 8.0, 24, and 70 $\mu$m.  The 24 $\mu$m band shows
a prominent point-like source at the central WD, superposed 
on an extended region of diffuse emission.  
A {\it Spitzer} IRS spectrum of the diffuse emission region shows 
that the [O\,IV] 25.89 $\mu$m line is the major contributor to 
the flux in the 24 $\mu$m band (see Figure 3a).  No IRS spectrum 
of the central point-like source is available.

\begin{figure}[pht]
\centerline{
\psfig{file=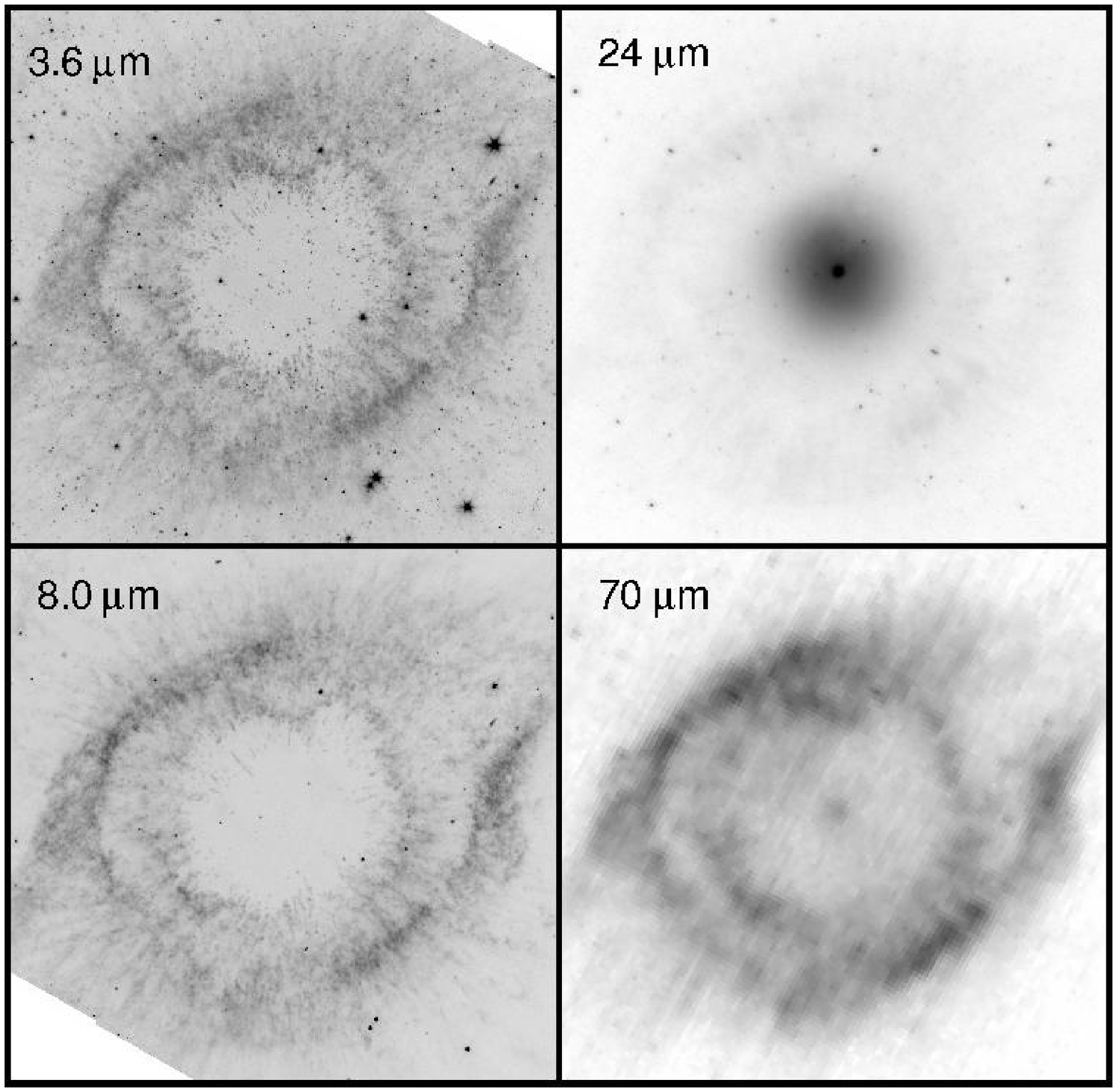,width=4.5in,angle=0}
}
\caption{
{\it Spitzer Space Telescope} images of the Helix Nebula.
}
\vspace*{1cm}
\centerline{
\psfig{file=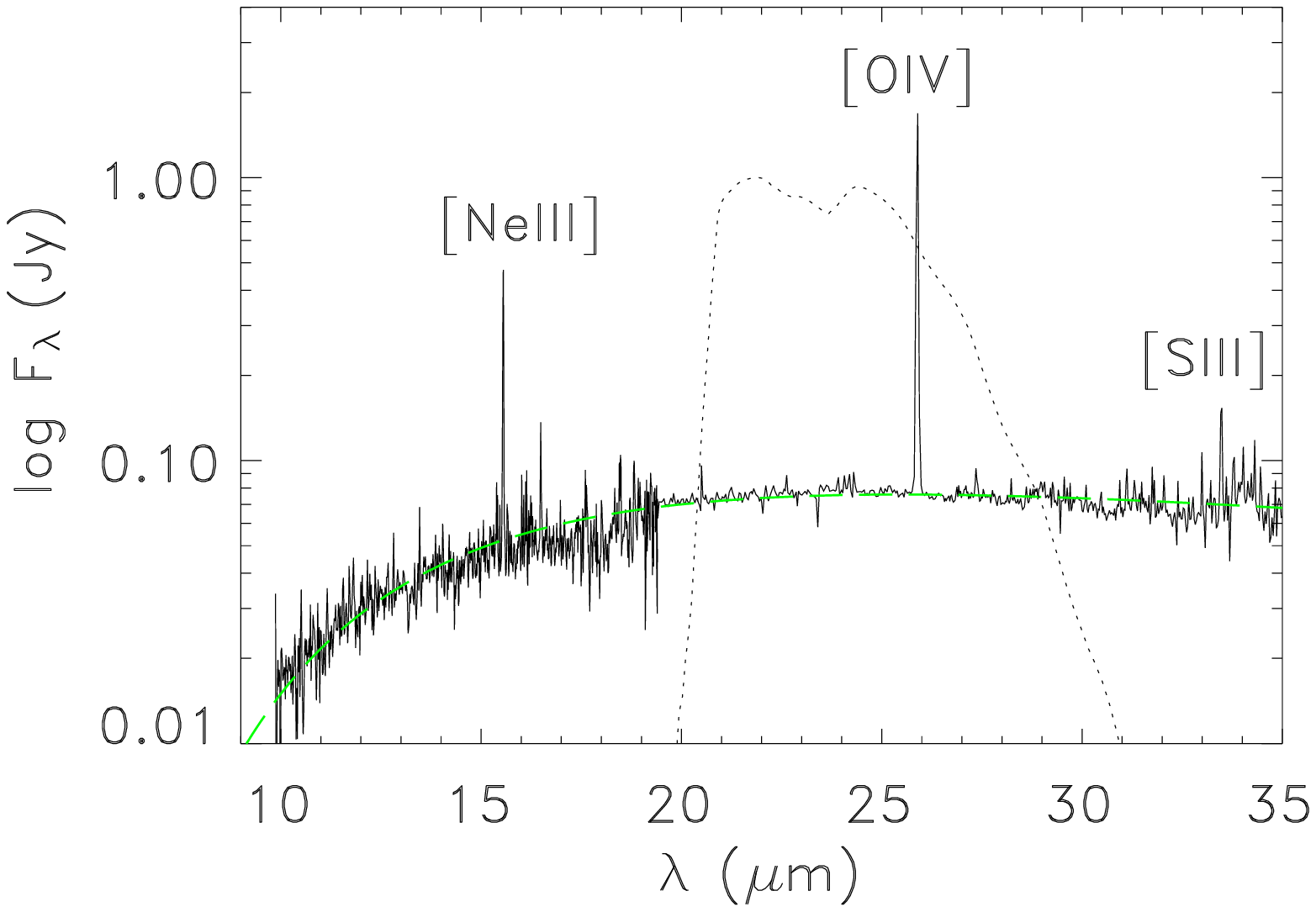,width=2.4in,angle=0}
\psfig{file=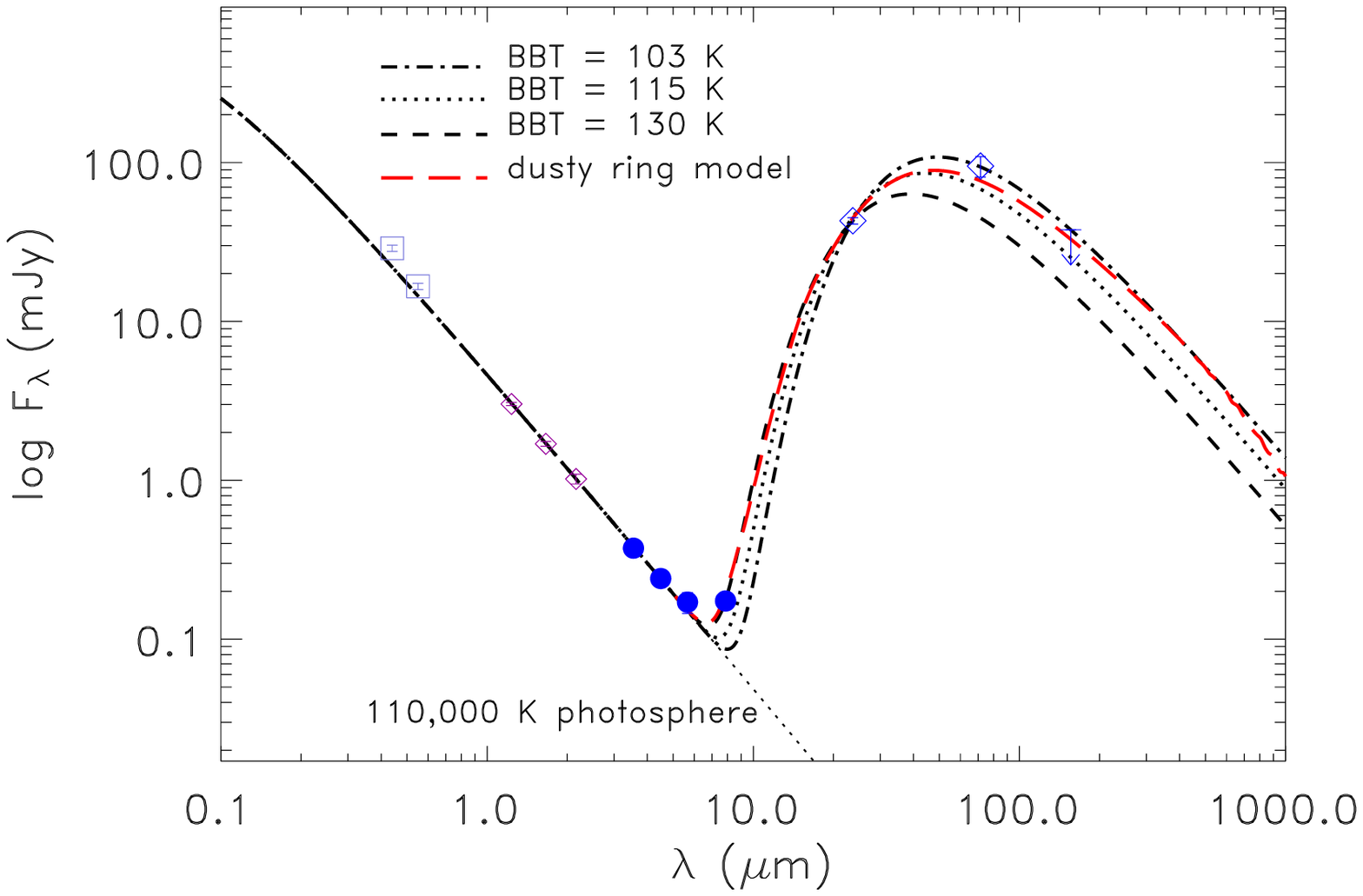,width=2.4in,angle=0}
}
\caption{
(a) {\it Spitzer} IRS spectrum of the central diffuse emission region
of the Helix Nebula.  The dotted curve shows the MIPS band.  (b) The
spectral energy distribution of the central point-like source that is
coincident with WD\,2226$-$210.  Three blackbody models and a dusty ring 
model are overplotted.
}
\end{figure}

To determine whether the central point-like source is also dominated
by the [O\,IV] 25.89 $\mu$m line emission in the 24 $\mu$m band, we
compare the 24 $\mu$m image with a He\,II $\lambda$4686 image.
The excitation potential of O\,IV, 54.9 eV, is similar to the 
ionization potential of He\,II, 54.4 eV; therefore, regions emitting
[O\,IV] 25.89 $\mu$m line should emit He\,II $\lambda$4686 line, too.
The He\,II image of the Helix Nebula \citep{odell98} shows a 
central diffuse emission region similar to that seen at 24 $\mu$m, 
but not a point-like central source.  Thus, we conclude that the 
[O\,IV] line does not contribute to the central point-like
source in the 24 $\mu$m band.  It is possible that the central 
source is dominated by the [Ne\,V] 24.32 $\mu$m line in the 
24 $\mu$m band, then the emission in the 70 $\mu$m band would
have to be dominated by high-ionization lines, but the 70 $\mu$m 
band contains only low-ionization lines.
Thus, it is most likely that the central point-like source 
detected in the MIPS 24 and 70 $\mu$m bands is a continuum 
source.

The spectral energy distribution (SED) of WD\,2226$-$210 is
presented in Figure 3b.  The optical and near-IR emission
follows the Rayleigh-Jeans tail, but the mid-IR emission 
rises sharply at 24 $\mu$m and remains high at 70 $\mu$m. 
This mid-infrared excess is consistent with a 100--130 K
blackbody emitter.  No stars have such a low temperature. 
If this IR excess emission is re-processed stellar emission,
the emitting material must be several tens of AU from 
WD\,2226$-$210, which has a stellar effective temperature of
$\sim$110,000 K.  For a distance of 210~pc \citep{Hetal97},
the 24 $\mu$m band luminosity, $\sim9\times10^{30}$ ergs~s$^{-1}$, 
requires an emitting area of almost 10 AU$^2$.
The temperature and distribution of the emitting material 
suggest that it is a dust disk.

Dusk disks have been observed around WDs, such as GD\,362 and 
G\,29-38 \citep{Betal05,Ketal05,ZB87,Retal05}, but these dust disks
are at smaller distances from their central WDs and have higher
temperatures.

\section{Hard and Mid-Infrared Emission from Apparently Single WDs}

Hard X-ray emission from the three apparently single WDs,
KPD\,0005+5106, PG\,1159, and WD\,2226$-$210, is puzzling.
The discovery of a dust disk around WD\,2226$-$210 opens up
the possibility that the accretion of disk material gives
rise to the hard X-ray emission.  The detailed process is
still unknown.  High-quality X-ray observations of KPD\,0005+5106
and PG\,1159 can provide spectral and temporal properties of
their hard X-rays and shed light on the origin of the X-ray
emission.  Mid-IR observations of these two WDs are also
needed to see whether they also have a dust disk.



\end{document}